# Calculation and Verification of Irradiation Damage Cross Section with Energy-Angular Distribution

Shengli Chen[1,2,*], David Bernard[1], Pierre Tamagno[1], and Cyrille De Saint Jean[1]

[1] CEA, Cadarache, DEN/DER/SPRC/LEPh, 13108 Saint Paul Les Durance, France

[2] Université Grenoble Alpes, I-MEP2, 38402 Saint Martin d'Hères, France

[*] Corresponding author: shengli.chen@cea.fr

**Abstract**

To complete the computation of Displacements per Atom (DPA) cross sections, the present work shows the methods of calculating DPA cross sections with the nuclear data of energy-angular distribution in both the laboratory and the Center-of-Mass (CM) frames. The method of direct calculation with data in the CM frame is proposed and recommended to decrease the computation burden and keep all information. Theoretical analyses reveal that more than 7-point Gauss-Legendre Quadrature (GLQ) should be used to ensure the convergence of the angular integration for DPA computations. Numerical results show that 8-point GLQ is sufficient for the continuum inelastic neutron scattering, while 64-point GLQ is implemented in NJOY. Because the integrand over secondary energy is not derivable in the whole domain of the secondary energy, the trapezoidal integration is used to perform the numerical integration. The numerical calculations show that the trapezoidal integration is suitable to perform the integration over the secondary energy on the fine grid given by nuclear data files at least for $^{56}$Fe. The present work reveals that the direct interpolation of energy-angular-integrated damage can give the same results computed with standard interpolated energy-angular distributions. The DPA cross sections will be overestimated if isotropic angular distributions are assumed. However, the first-order Legendre polynomial can give DPA cross sections within 0.4% deviation, while 12 orders are required to describe the anisotropic angular distribution.

**Keywords:** DPA, cross section, energy-angular distribution, Gauss-Legendre quadrature, trapezoidal integration, $^{56}$Fe

## 1. Introduction

The Displacement per Atom (DPA) is conventionally used to quantify the irradiation damage of materials. In nuclear industry, the neutron embrittlement is one of the three major materials challenges of the RPV [1]. It is thus of importance to accurately compute the neutron-induced DPA. The DPA is calculated using the energy of Primary Knock-on Atom (PKA), for both Molecular Dynamics (MD) simulations, Binary Collision Approximation (BCA), and the standard metrics summarized in Section 2 in the previous manuscript [2]. The neutron-induced DPA is calculated by determining the recoil energy of target nuclei after nuclear reactions with neutrons.

The previous work [2] studied the DPA cross sections by using the angular distribution given in Evaluated Nuclear Data Files (ENDF). A more suitable method of angular integration has been proposed to ensure the convergence of DPA cross sections. However, due to the undetermined reaction $Q$-value for several nuclear reactions, such as the continuum inelastic scattering, one more degree of freedom should be considered because the conservation of energy cannot reduce one unknown parameter. The DPA cross sections are calculated with a double integration over the emission angle and the secondary energy. In addition, the interpolation of energy-angular distribution using the tabulated data given in ENDF should be performed on two dimensions (incident energy and secondary energy), while only the linear interpolation over incident energy is required for the angular distribution studied in the previous work [2]. The present work completes the investigation of neutron-induced DPA cross section with energy-angular distribution given in ENDF.

The computation of DPA cross sections in the Laboratory (Lab) frame was well developed (Section 2.1) and used in NJOY [3]. Three methods to calculate DPA with energy-angular distributions given in the Center-of-Mass (CM) frame are presented and compared in Section 2.2. The methods and reliability of integration for continuum energy-angular distributions are investigated in Sections 3.1 and 3.2. Because the energy-angular distributions are tabulated on coarse meshes of incident energy given in ENDF, Section 3.3 shows the method to compute the DPA cross sections between two incident energies.

Refs. [2], [4], [5] show the influence of the anisotropic angular distribution on PKA energy. The previous work reveals that the high-order Legendre polynomials are not important in the calculation of DPA for elastic and discrete inelastic scatterings [2]. The influence of high-order Legendre polynomials of angular distribution is investigated in Section 3.4 for neutron continuum inelastic scattering. As explained in the previous manuscript [2], $^{56}$Fe is studied because the Stainless Steel (SS) is used for the RPV in LWR, candidate fuel cladding in Accident Tolerant Fuel (ATF) of LWR [6], and fuel cladding in Fast Reactor (FR). The numerical results shown in the following studies are based on the JEFF-3.1.1 nuclear data library [7].

## 2. Methods

2.1 DPA cross sections with the energy-angle distribution in the laboratory frame

As mentioned in Section 1, the recoil energy of PKA can be computed with the energy-angular distribution of the secondary particle. Figure 1 shows the schemes of the collision in the Lab and CM frames. The incident and emitted energies (velocities) are referred respectively to $E$ and $E'$ ($v_0$ and $v'$) in the Lab frame. $E_R$ ($v_R$) stands for the recoil energy (velocity) of the target in the Lab frame. $m$ and $v_1$ ($m'$ and $u_1$) are respectively the mass and velocity of incident (outgoing) particle in the CM frame. $M$ and $v_2$ ($M'$ and $u_2$) are respectively the mass and velocity of the target particle before (after) the collision in the CM frame.

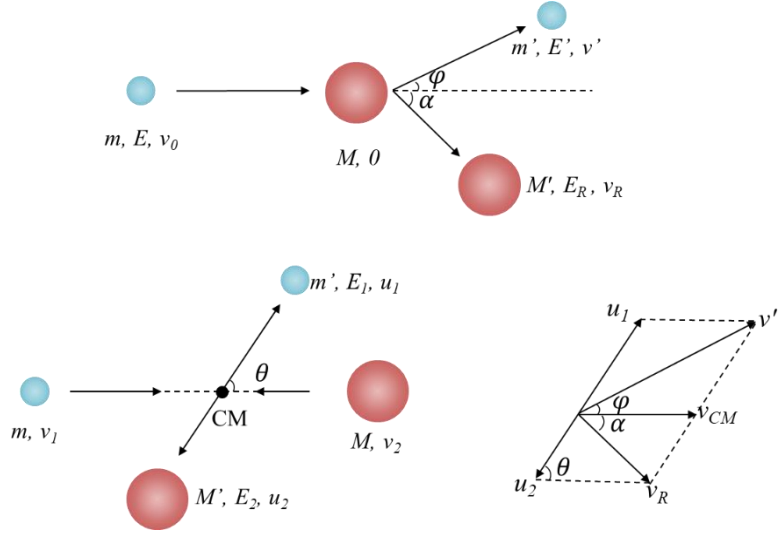

Figure 1. Schemes of the collision in the Laboratory (upper) and Center-of-Mass (lower) frames

The relativity effect can be neglected in our studies that focus on energies lower than 20 MeV [8]. The conservation of momentum in the Lab frame points out:

$$mv_0 = m'v'\cos\varphi + M'v_R\cos\alpha, \quad (1)$$

$$m'v'\sin\varphi = M'v_R\sin\alpha. \quad (2)$$

By eliminating $\alpha$, the recoil energy of PKA is obtained as:

$$E_R(E, E', \tilde{\mu}) = \frac{1}{M'}[mE - 2\sqrt{mm'EE'}\tilde{\mu} + m'E']. \quad (3)$$

where $\tilde{\mu} = \cos\varphi$.

The Robinson's damage energy is given by [9]:

$$E_a(E, E', \tilde{\mu}) = E_R(E, E', \tilde{\mu})P(E_R(E, E', \tilde{\mu})/E_L). \quad (4)$$

The energy-angular-integrated DPA cross section related to a given reaction is obtained by:

$$\sigma_{DPA}(E) = \sigma(E)\int_0^\infty \int_{-1}^1 \tilde{f}(E, E', \tilde{\mu})E_a(E, E', \tilde{\mu})\xi(E_a)d\tilde{\mu}dE', \quad (5)$$

where $\sigma(E)$ is the corresponding reaction cross section, $\xi(E_a)$ is the efficiency of displacement based on the NRT metric, it is unity for the NRT-DPA and Eq. (5) in Ref. [2] for the ARC-DPA model. $\tilde{f}(E, E', \tilde{\mu})$ is the probability density of energy-angular distribution in the Lab frame for the incident energy $E$ versus the secondary energy $E'$ and the cosine of the emission angle $\tilde{\mu}$. $\tilde{f}(\tilde{\mu}, E, E')$ is conventionally given by the combination of Legendre polynomials:

$$\tilde{f}(E, E', \tilde{\mu}) = \sum_{l=0}^{Lmax} \frac{2l+1}{2}\tilde{b}_l(E, E')P_l(\tilde{\mu}), \quad (6)$$

where $P_l$ is the $l$-th Legendre polynomial and $\tilde{b}_l(E, E')$ is the corresponding Legendre coefficient given in Evaluated Nuclear Data Files (ENDF).

2.2 DPA cross sections with the energy-angle distribution in the center of mass frame

The energy-angle distributions are suggested to be given in the Lab frame in ENDF. However, they are often provided in the CM frame on which many nuclear theories are based, including JEFF-3.1.1 [7]. Three methods can be used to compute the DPA cross section with double-differential cross section given in the CM frame.

*2.2.1 Relationship between variables in the CM frame and the Lab frame*

**Lab to CM:** The velocity of the outgoing particle in the CM frame is:

$$u_1^2 = v'^2 + v_{CM}^2 + 2v'v_{CM}\tilde{\mu}, \qquad (7)$$

where the velocity of the center of mass is determined by the conservation of momentum:

$$(m+M)v_{CM} = mv_0, \qquad (8)$$

where $v_0$ and $v_{CM}$ are respectively the initial velocity of the incident particle and the velocity of the Center of Mass in the Lab frame. The explicit expression of $E_1$ with the quantities in the Lab frame is thus:

$$E_1 = E' + \frac{mm'E}{(m+M)^2} - 2\sqrt{\frac{mm'EE'}{(m+M)^2}}\tilde{\mu}. \qquad (9)$$

Projecting the velocity into the incident direction (see the lower right scheme in Figure 1) leads to:

$$u_1\mu + v_{CM} = v'\tilde{\mu}. \qquad (10)$$

where $\mu = cos\theta$ is the cosine of the emission angle in the CM frame. The explicit expression of $\mu$ knowing $\tilde{\mu}$ and $E'$ can be deduced from Eqs. (8) and (10):

$$\mu = \sqrt{\frac{E'}{E_1}}\left(\tilde{\mu} - \sqrt{\frac{mm'E}{(m+M)^2 E'}}\right). \qquad (11)$$

**CM to Lab:** Transforming the recoil velocity of the emitted particle from the CM to the Lab frame (see the lower right scheme in Figure 1):

$$v'^2 = u_1^2 + v_{CM}^2 + 2u_1 v_{CM}\mu, \qquad (12)$$

Consequently, the secondary energy in the Lab frame can be obtained with Eqs. (8) and (12):

$$E' = \frac{mm'E}{(m+M)^2} + E_1 + 2\frac{\sqrt{mm'EE_1}}{m+M}\mu, \qquad (13)$$

On the other hand, projecting the velocity in Eq. (10) implies:

$$\tilde{\mu} = \frac{u_1\mu + v_{CM}}{v'}. \qquad (14)$$

One can obtain further the expression:

$$\tilde{\mu} = \frac{\sqrt{mm'E}+(m+M)\sqrt{E_1}\mu}{\sqrt{(m+M)^2 E_1 + mm'E + 2(m+M)\sqrt{mm'EE_1}\mu}}. \quad (15)$$

For neutron scatterings, it can be simplified to:

$$\tilde{\mu} = \frac{\sqrt{E}+(1+A)\sqrt{E_1}\mu}{\sqrt{(1+A)^2 E_1 + E + 2(1+A)\sqrt{EE_1}\mu}}, \quad (16)$$

where $A$ is the relative mass to neutron of the target nucleus.

*2.2.2 Transformation of data from the CM frame to the Lab frame*

Section 2.1 shows the routine of DPA calculations with energy-angular distribution in the Lab frame. For the data given in the CM frame, this method can be applied by transforming the data in the CM frame to the Lab frame. The transformation of data from the CM frame to the Lab frame is also the strategy of NJOY [3]. For a given incident energy $E$, the coefficients $\tilde{b}_l(E, E')$ in Eq. (6) can be determined through the energy-angular distribution provided in the CM frame $f(E, E_1, \mu)$, in which $\mu = cos\theta$ and $E_1$ is the secondary energy in the CM frame. This method is implemented in NJOY because the Legendre coefficients in the Lab frame can be used to compute all corresponding quantities in the same frame.

Because the Legendre polynomials are orthogonal (and orthonormal for $((2l+1)/2)^{1/2} P_l$) with respect to the $L^2$ norm on the interval [-1,1], the coefficients $\tilde{b}_l(E, E')$ can be calculated by:

$$\tilde{b}_l(E, E') = \int_{-1}^{1} \tilde{f}(E, E', \tilde{\mu}) P_l(\tilde{\mu}) d\tilde{\mu}. \quad (17)$$

Since there are 2 degrees of freedom for the energy-angular distribution, the passage from the CM frame to the Lab frame should be performed with double integrals.

$$\int_{-1}^{1} \tilde{f}(E, E', \tilde{\mu}) P_l(\tilde{\mu}) d\tilde{\mu} = \int_{-1}^{1} \int_{0}^{E'_{max}} \delta_{E'}(E'') \tilde{f}(E, E'', \tilde{\mu}) P_l(\tilde{\mu}) dE'' d\tilde{\mu}, \quad (18)$$

where $E'_{max}$ can be determined by Eq. (13) and the Dirac delta function about $E'$ is defined as:

$$\delta_{E'}(E'') = \begin{bmatrix} 1, & E'' = E' \\ 0, & \text{otherwise} \end{bmatrix}. \quad (19)$$

By using the data in the CM frame, the Legendre coefficients in the Lab frame are:

$$\tilde{b}_l(E, E') = \int_{-1}^{1} \int_{0}^{E_{1,max}} \delta_{E'}(E_1, \mu) f(E, E_1, \mu) P_l(\tilde{\mu}(E_1, \mu)) dE_1 d\mu, \quad (20)$$

where the Dirac delta function links two variables in the CM frame with $E'$:

$$\delta_{E'}(E_1, \mu) = \begin{bmatrix} 1, & \text{Eq. (13)}: (E_1, \mu) \to E' \\ 0, & \text{otherwise} \end{bmatrix}. \quad (21)$$

The maximum secondary energy in the CM frame ($E_{1,max}$) in Eq. (20) is given by ENDF.

The change of variables in double integrals for Eq. (20) conducts to:

$$\tilde{b}_l(E, E') = \int_{\tilde{\mu}_{min}}^{1} f(E, E_1(E', \tilde{\mu}), \mu(E', \tilde{\mu})) P_l(\tilde{\mu}) J(E) d\tilde{\mu}. \quad (22)$$

where the Jacobian $J(E)$ is the determinant of the Jacobian matrix of the transformation from $(E', \tilde{\mu})$ to $(E_1, \mu)$:

$$Jac(E, E', \tilde{\mu}) = \begin{bmatrix} \partial E_1/\partial E' & \partial E_1/\partial \tilde{\mu} \\ \partial \mu/\partial E' & \partial \mu/\partial \tilde{\mu} \end{bmatrix}. \quad (23)$$

The determinant of the Jacobian matrix calculated with Eqs. (9), (11), (13), and (23) is:

$$J(E) \equiv \det[Jac(E, E_1, \mu)] = \sqrt{\frac{E_1}{E'}}. \quad (24)$$

$J(E)$ rather than $|J(E)|$ is used in Eq. (12) because the Jacobian is always positive, shown with Eq. (14).

The lower limit of the integral in Eq. (22) is not necessarily -1 because the minimum value of $\tilde{\mu}$ for a given $E'$ can be larger than -1. This is due to the limits in [-1,1] for $\mu(E', \tilde{\mu})$. According to Eq. (15), the lower limit of the integration in Eq. (22) is:

$$\tilde{\mu}_{min}(E_{1,max}) = \max\left\{\frac{\sqrt{mm'E}}{(m+M)\sqrt{E'}} - \sqrt{\frac{E_{1,max}}{E'}}, -1\right\}, \quad (25)$$

For a given $(E, E')$, $E_1$ is a function of $\tilde{\mu}$. Calculation of $\tilde{b}_l(E, E')$ by Eq. (22) requires the density $f(E, E_1(E', \tilde{\mu}), \mu(E', \tilde{\mu}))$ for each $\tilde{\mu}$. The energy-angular distributions are usually tabulated for the secondary energy $E_1$. The interpolation of $f(E, E_1, \mu)$ on the secondary energy grid is required for each $\tilde{\mu}$. This method increases the computation burden.

Moreover, for a given incident energy, we should define the suitable grid of the secondary energies $(E')$. If the grid is too fine, too many calculations are required. If the grid is too coarse, some information will be lost. NJOY takes the criterion that the difference between the coefficient of the midpoint in each interval calculated by Eq. (22) and the linearly interpolated value with two boundaries should be less than 2% [10]. Anyway, transforming the data of energy-angular distribution in the CM frame to the Lab frame gives an additional error for DPA cross section.

*2.2.3 Change of variables in double integrals*

The change of variables is an intuitive method for the transformation of frames. This method can avoid the problem of the loss of information. Because the Jacobian is always positive, the change of double variables in the CM frame to the Lab frame leads to:

$$\sigma_{DPA}(E) = \sigma(E) \int_0^\infty \int_{-1}^1 F(E, E_1, \mu) [J(E)]^{-1} d\mu dE_1, \quad (26)$$

where the Jacobian $J(E)$ is found in Eq. (24) with $E'$ in Eq. (13), and

$$F(E, E_1, \mu) = f(E, E_1, \mu) E_a(E, E'(E_1, \mu), \tilde{\mu}(E_1, \mu)) \xi(E_a(E_1, \mu)), \quad (27)$$

where $E'(E_1, \mu)$ and $\tilde{\mu}(E_1, \mu)$ are given in Eqs. (13) and (15), respectively.

The DPA cross sections with the energy-angular distributions provided in the CM frame can be computed with Eqs. (26) and (27). All information given in the CM frame can be used in the computation of DPA cross sections. However, the integrand in Eq.

(5) has a much simpler form than the integrand in Eq. (26). A consequent result is that the numerical integration for Eq. (26) converges more slowly than Eq. (5). In other words, comparing with the energy-angular distribution given in the Lab frame and DPA cross sections computed with Eq. (5), finer grids are required to perform the numerical integrals of Eq. (26) in the case of double-differential nuclear data given in the CM frame.

*2.2.4 Calculation in the CM frame*

The above methods can compute the DPA cross sections with the energy-angular distribution in the CM frame. The transformation of data between two frames increases the computation burden and introduces additional error. The change of variables is more feasible than the transformation of frames for the calculations of DPA cross section. However, comparing with the double-differential nuclear data given in the Lab frame and the DPA cross sections calculated by Eq. (5), the change of variables increases the computation burden.

In fact, the direct calculation of DPA cross sections in the CM frame is much simpler than the two previous methods for the energy-angular distributions given in the same frame. The momentum in the CM frame is always null. The conservation of momentum points out:

$$m'u_1 = M'u_2. \quad (28)$$

Transforming the recoil velocity from the CM frame to the Lab frame reveals:

$$v_R^2 = u_2^2 + v_{CM}^2 - 2u_2 v_{CM}\mu. \quad (29)$$

Consequently, the recoil energy can be obtained with the above equations:

$$E_R(E, E_1, \mu) = \frac{mM'}{(m+M)^2}E - 2\frac{\sqrt{mm'EE_1}}{m+M}\mu + \frac{m'}{M'}E_1, \quad (30)$$

The energy-angular-integrated DPA cross section can be directly computed with:

$$\sigma_{DPA}(E) = \sigma(E)\int_0^\infty \int_{-1}^1 f(E, E_1, \mu)\, E_a(E, E_1, \mu)\xi(E_a)d\mu dE_1, \quad (31)$$

where $f(E, E_1, \mu)$ is the probability density of energy-angular distribution in the CM frame for the incident energy $E$ versus the secondary energy $E_1$ and the cosine of the emitted angle $\mu$. $f(\mu, E, E_1)$ is conventionally recommended with a combination of Legendre polynomials:

$$f(E, E_1, \mu) = \sum_{l=0}^{Lmax} \frac{2l+1}{2} b_l(E, E_1) P_l(\mu), \quad (32)$$

where $P_l$ is the $l$-th Legendre polynomial and $b_l(E, E_1)$ is the corresponding Legendre coefficient in the CM frame given in ENDF.

Comparing with the change of variables, this method simplifies the calculations. In addition, the integrand in Eq. (31) has a simpler form than that in Eq. (28). Therefore, numerical methods converge more quickly for the direct calculation in the CM frame than the change of variables. As a matter of fact, Eqs. (5) and (31) have the similar form, the computation of DPA cross sections (by Eq. (31)) with the energy-angular distributions given in the CM frame converges as quickly as the calculation (by Eq. (5)) with double-differential data provided in the Lab frame. Therefore, the method of direct computation in the CM frame with Eq. (31) is recommended if the energy-angular distributions are given in the CM frame.

2.3 Minimum recoil energy

The DPA cross section shown in Eqs. (5), (26), and (31) are available only for $E_a > 2E_d/0.8$ when the DPA metrics such as NRT and ARC are used. In order to generalize those expressions of DPA cross section in the interval $[0, 2E_d/0.8]$, one should define the generalized damage energy as:

$$\widetilde{E_a} = \begin{bmatrix} 0, & 0 < E_a < E_d \\ 2E_d/0.8, & E_d < E_a < 2E_d/0.8 \\ E_a, & 2E_d/0.8 < E_a < \infty \end{bmatrix}. \quad (33)$$

This generalization implies that the damage energy is not a continuous function. More precisely, it is not continuous at $E_d$, continuous but not derivable at $2E_d/0.8$. A Gauss-Legendre Quadrature-based Piecewise Integration (GLQPI) method has been proposed to perform the integral of such a non-continuous damage energy over the whole interval of cosine to ensure the convergence [2]. The objective of this sub-section is to show that the normal GLQ can be used in the whole domain of emission angle and secondary energy for the continuum inelastic scattering.

For the continuum energy-angle distributions, the minimum recoil energy is:

$$\min_\mu E_R(E, E_1, \mu) = \left( \frac{\sqrt{mM'E}}{m+M} - \sqrt{\frac{m'E_1}{M'}} \right)^2. \quad (34)$$

On the other hand, the conservation of energy in the CM frame implies that:

$$(m' + M')c^2 + E_1 + E_2 + Q'(E) = (m + M)c^2 + E - \frac{1}{2}(m + M)v_{CM}^2. \quad (35)$$

where $Q'(E)$ is the loss of energy due to inelastic scattering. The dependence on energy is due to the unresolved excitation level of the knocked-on nucleus. This equation induces:

$$E_1 = \frac{MM'}{(m+M)(m'+M')}E - \frac{M'}{(m'+M')}Q(E), \quad (36)$$

where $Q(E) = Q'(E) + [(m' + M') - (m + M)]c^2$ is the total loss of energy, which is then transferred to the excitation energy of the target nucleus. Usually, $Q \geq 0$ because the total energy of the isolated system cannot increase. $Q = 0$ if and only if the reaction is elastic scattering. Eq. (36) implies:

$$\frac{\sqrt{mM'E}}{m+M} / \sqrt{\frac{m'E_1}{M'}} \geq \sqrt{\frac{mM'}{m'M} \frac{1}{1-(m+M)/(MQ_{min}/E)}}, \quad (37)$$

where the minimum value of energy loss ($Q_{min}$) is given in ENDF.

The right-side term in Eq. (37) is always larger than unity for scattering reactions because $mM' = m'M$. The minimum recoil energy of scattering reactions for the incident energy $E$ is thus:

$$E_{R,min}(E) = \left(\frac{\sqrt{mM'E}}{m+M} - \sqrt{\frac{m'ME}{(m+M)(m'+M')} - \frac{m'Q_{min}}{(m'+M')}}\right)^2, \qquad (38)$$

or formed as:

$$E_{R,min}(E) = \left(\frac{\sqrt{m}Q_{min}}{\sqrt{ME}+\sqrt{ME-(m+M)Q_{min}}}\right)^2, \qquad (39)$$

which is a decreasing function of the incident energy $E$. In fact,

$$E_{R,min}(E) > \frac{mQ_{min}^2}{4ME}, \qquad (40)$$

is quite large compared with the threshold energy of displacement. The minimum recoil energy should be larger than $2E_d/0.8$ in the cases of the applications in nuclear reactors, i.e. lower than 20 MeV.

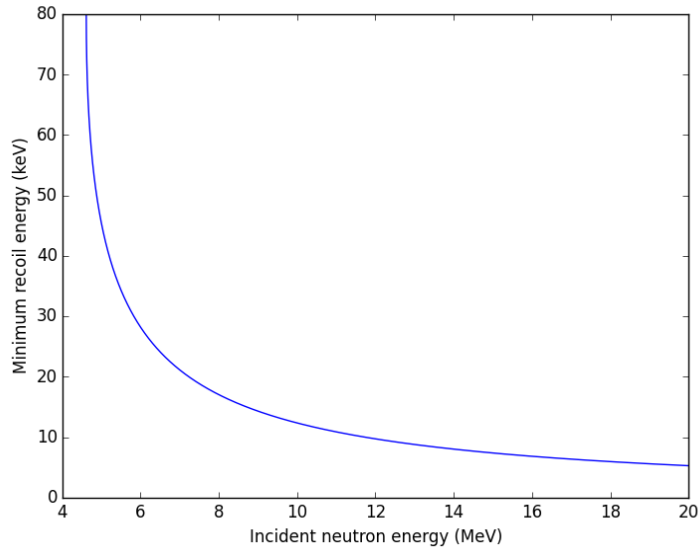

Figure 2. Minimum recoil energy of the continuum inelastic scattering of $^{56}$Fe (computed with Eq. (39)) based on JEFF-3.1.1.

The example of the minimum recoil energy computed with Eq. (39) for the continuum inelastic scattering of $^{56}$Fe is shown in Figure 2. The minimum recoil energy is 5 keV at 20 MeV incident neutron energy, while $2E_d/0.8 = 100$ eV. Consequently, the recoil energy is a smooth function of the emission angle and secondary energy for continuum inelastic scattering, so is the damage energy. The Gauss-Legendre Quadrature (GLQ) can be used to compute angle-integrated DPA cross sections in the whole interval with high accuracy.

## 3. Results and Discussion

### 3.1 Angular integration

The integral of the DPA cross sections over the emission angle is performed with a 4-point GLQ in NJOY if the energy-angular distribution is given in file MF5 because of the assumption of quasi-isotropic angular distribution [3]. A 64-point GLQ is used by NJOY when the energy-angular distribution is given in file MF6. The angular distributions are not isotropic, especially for high secondary energies. Figure 3 shows the angular distributions of the 14 MeV continuum inelastic scattering of $^{56}$Fe with different secondary energies. The strong forward oriented distribution for high emitted energy can be found. The product of the recoil energy and probability density is 13-order polynomials. 7-point GLQ can exactly compute the integration of 13-order polynomials. Due to the product with the partition function, which is not a polynomial of the cosine, more than 7 points should be used to compute the DPA cross sections with the GLQ for angular integration.

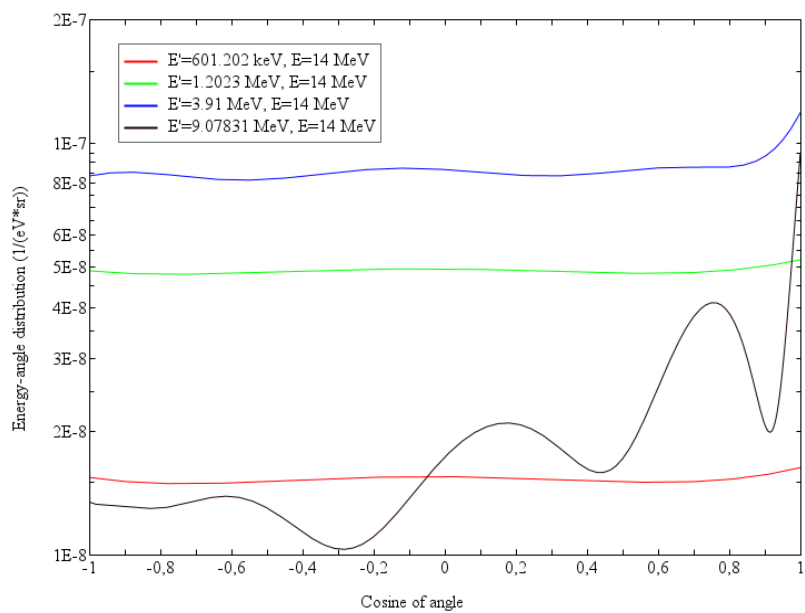

Figure 3. Angular distribution of the 14 MeV continuum inelastic scattering of $^{56}$Fe in JEFF-3.1.1 for different exit energies. Both the cosine and the secondary energies are data in the CM frame.

However, 64 points are too many because 20 points are shown sufficient for the elastic and discrete inelastic scatterings, for which 19 orders of Legendre polynomials are used [2]. For the continuum inelastic scattering with 12 orders, less than 20 points should be enough to perform the integral with the GLQ method. In addition, the DPA cross section is the double integral over angular distribution and secondary energies. Due to the contribution of quasi-isotropic angular distributions that require less Legendre polynomials for some secondary energies (such as 1.2 MeV in Figure 3), the angular integration should converge quickly with the number of points used in the GLQ.
10

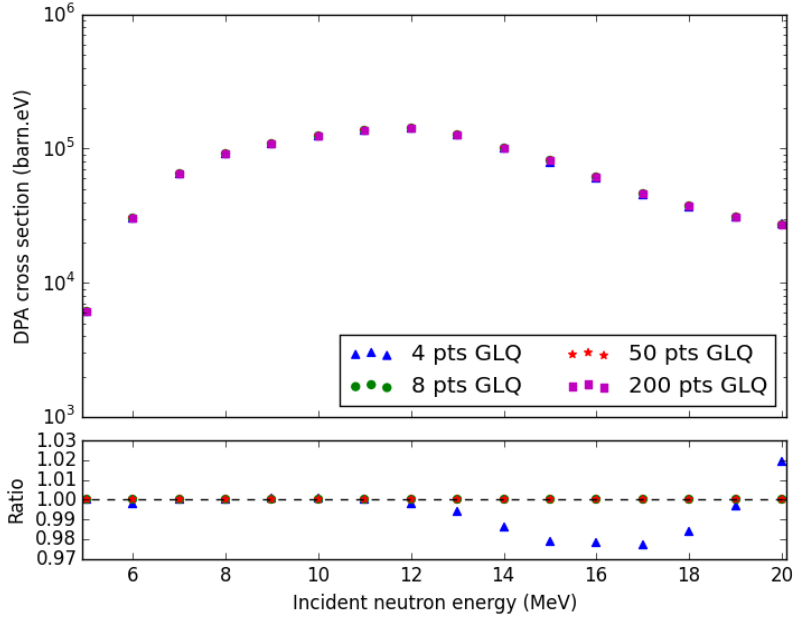

Figure 4. Continuum inelastic scattering DPA cross sections of $^{56}$Fe performed with 4-, 8-, 50- and 200-point GLQ and the corresponding ratios to the 200-point GLQ calculations.

Figure 4 shows the convergence of the integration performed with 4-, 8-, 50- and 200-point GLQ and the corresponding ratios to the 200-point GLQ results. Only the data with the energy-angular distributions explicitly given in JEFF-3.1.1 are shown. As explained in the above paragraph, the 4-point GLQ is not sufficient to ensure the convergence of the angular integration. Potential 3% deviation can be found for the 4-point GLQ angular integration. As expected previously, the DPA cross sections converge quickly with the number of points used in the angular integration. 8-point GLQ is shown sufficient for the angular integration.

3.2 Integration over the secondary energy

For the elastic and discrete inelastic scatterings, the DPA cross sections are computed with only the angular integration because the secondary energy is determined by the emission angle and the constant *Q*-value of the reaction. In other words, the elastic and discrete inelastic scatterings have only 1 degree of freedom. For the continuum inelastic scatterings, the integration over the secondary energy should be computed because the emission angle and the secondary energy are two independent variables. The secondary energy distribution of the continuum inelastic scattering of $^{56}$Fe is plotted in Figure 5. For the tabulated secondary energy distribution, which is often the case, NJOY uses the trapezoidal integration with the energy grid in the ENDF. However, Eq. (3) or Eq. (30) indicates that the recoil energy is not a linear function of the secondary energy, neither is the angle-integrated damage energy. The ENDF energy grid-based trapezoidal integration is thus not necessarily accurate. The error of the trapezoidal integration on each interval is dominated by the second derivative of the integrand.

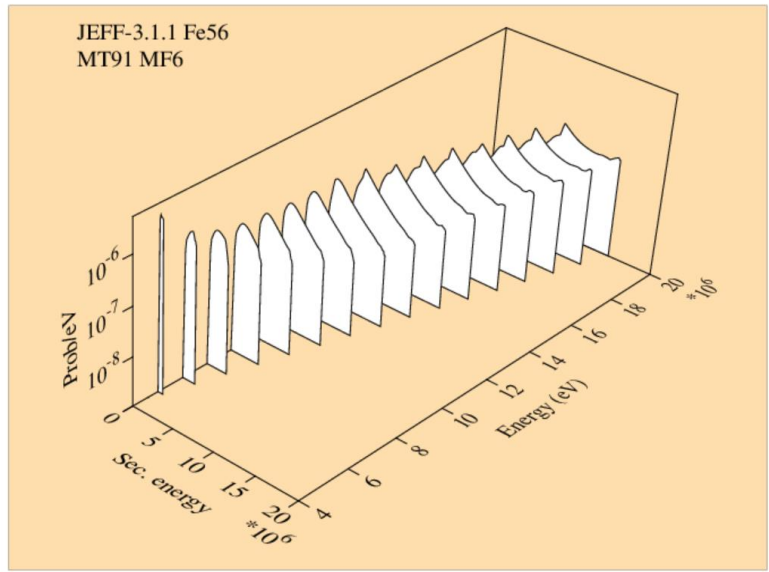

Figure 5. Energy distribution of the continuum inelastic scattering of $^{56}$Fe in JEFF-3.1.1

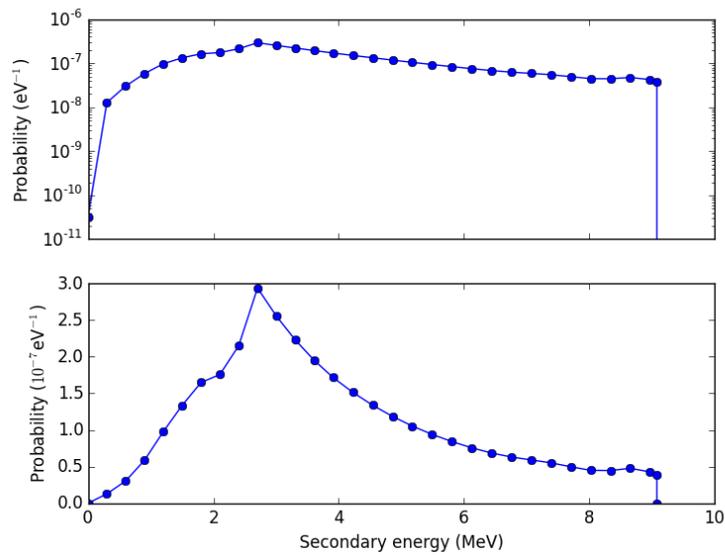

Figure 6. Energy distribution of 14 MeV continuum inelastic scattering of $^{56}$Fe $f(E = 14 \text{ MeV}, E_1, \mu = 0)$ in the CM frame (JEFF-3.1.1).

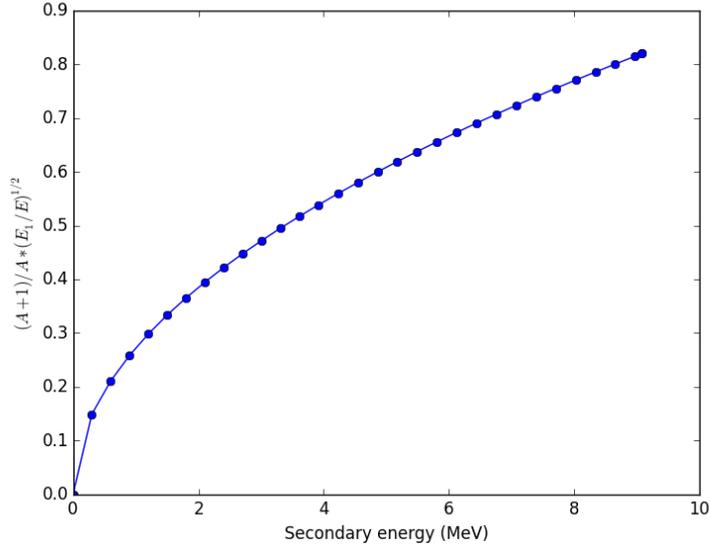

Figure 7. $\sqrt{\frac{m'E_1}{M'}} / \frac{\sqrt{mM'E}}{m+M}$ versus $E_1$ for 14 MeV continuum inelastic scattering of $^{56}$Fe. The points are calculated based on the energy grid in JEFF-3.1.1.

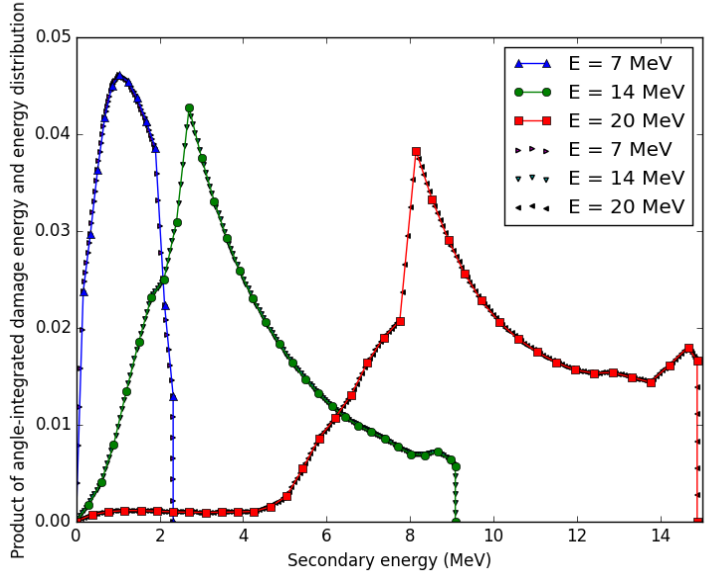

Figure 8. Energy distribution of the angle-integrated damage energy, i.e. $\int_{-1}^{1} f(E, E_1, \mu) E_a(E, E_1, \mu) d\mu$, for 7 MeV, 14 MeV, and 20 MeV neutron continuum inelastic scattering of $^{56}$Fe. The larger points are calculated based on the energy grid in JEFF-3.1.1. The smaller points are computed at 5 evenly inserted energies.

As mentioned in the above section, the angular integrations are performed with the GLQ. The integral over angles in the GLQ is given by a weighted combination of damage energy:

$$\int_{-1}^{1} f(E, E_1, \mu) E_a(E, E_1, \mu) d\mu = \sum_{i=1}^{N} w_i f(E, E_1, x_i) E_a(E, E_1, x_i), \qquad (41)$$

where $w_i$ and $x_i$ are respectively the *i*-th weight and the *i*-th Gauss node (*i*-th zero of $P_N$) in the *N*-point GLQ. The previous study shows that $N = 8$ is sufficient. The probability density function $f(E, E_1, x_i)$ is conventionally a linearly interpolated function between two secondary energies. The example of 14 MeV neutron inelastic scattering is shown in Figure 6.

The recoil PKA energy is not a polynomial function as the secondary energy due to the square root term (shown in Eq. (30)) as the example plotted in Figure 7. The partition function is never a polynomial function of neither the cosine of emission angle nor the secondary energy. The probability density has different Legendre polynomial expressions in each interval of the energy grid due to the linear interpolation between two tabulated points. The trapezoidal integration is one of the best methods to numerically compute the integral for the secondary energy distribution. The accuracy depends on the energy grid. The secondary energy grids in the ENDF-6 format nuclear data files, especially those of important isotopes, are normally fine because they are used to describe details of the energy-angular distribution.

The energy distribution of the angle-integrated damage energy of $^{56}$Fe is shown in Figure 8. The scattered points are computed with the data in JEFF-3.1.1. The dashed lines are the linear interpolation of scattered points. The trapezoidal integral is equivalent to the area of the zone enclosed by the dashed lines and *x*-axis. Thanks to the fine grid of secondary energy, the error of the trapezoidal integration should be small.

To verify the conclusion about the convergence of the trapezoidal integration in the calculations of DPA cross sections, additional equidistant points are inserted in each interval of the original energy grid, such as the 5-inserted points illustrated in Figure 9. The calculations performed with the 5 evenly inserted points are shown in Figure 8 with smaller triangles. As expected, the results in Figure 8 are quite similar to those in Figure 8 due to the quasi-linear damage energy versus the secondary energy and linear interpolation of angular distribution, which is a standard method to determine the energy-continuous angular distribution in nuclear data interpretation.

Figure 10 shows the DPA cross sections of the continuum inelastic scattering of $^{56}$Fe performed with the trapezoidal integration using the energy grid in JEFF-3.1.1, 5- and 50-inserted equidistant points in each interval of the secondary energy in JEFF-3.1.1. The numerical results show that the trapezoidal integration using the energy grid in ENDF can give accurate results, at least for $^{56}$Fe, while 2% potential error is permitted in NJOY during the transformation of data from the CM frame to the Lab frame [3].

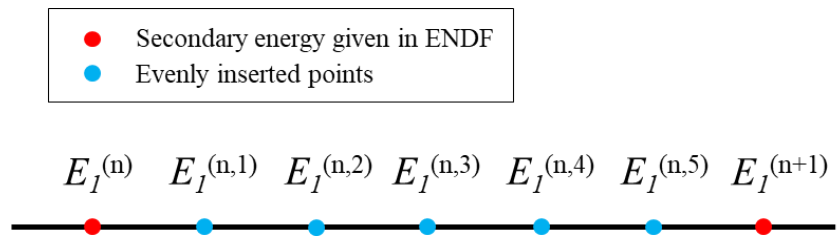

Figure 9. Scheme of 5 evenly inserted points between two neighbor secondary energies given in ENDF

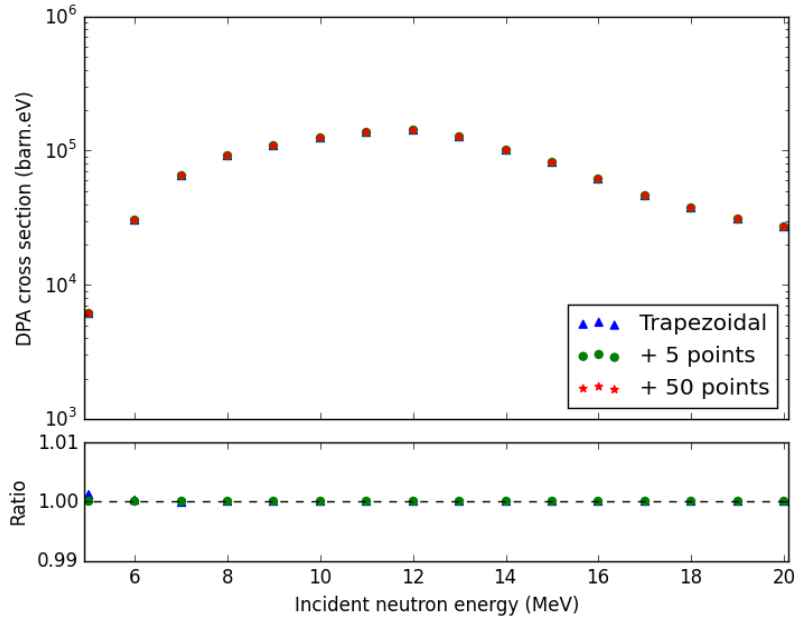

Figure 10. Neutron continuum inelastic scattering DPA cross sections of $^{56}$Fe computed with the trapezoidal integration by using the energy grid in ENDF, 5 and 50 evenly inserted points in each interval. The corresponding ratios are calculated over the DPA with 50 inserted points.

3.3 Computation of DPA cross sections between two incident energies

    The above analyses of DPA cross sections are based on the incident energy at which the energy-angular distribution is given in the ENDF. To compute the DPA cross sections between two neighbor incident energies, NJOY uses the linear interpolation of the energy-angular-integrated damage energy. In fact, the physical method is to compute DPA cross sections using the interpolated the energy-angular distribution. The The most common method Unit-Base Interpolation (UBI) [11] is used to interpolate the energy-angular distribution. In order to verify the direct interpolation of damage energy, the DPA cross sections computed with the interpolation of damage energy are compared with those calculated with interpolated energy-angular distribution.

    Figure 11 reveals the energy-angular-integrated damage energies at different incident energies. The red square points are computed with energy-angular distributions given in the ENDF, the red lines are linear interpolation between two incident energies, the blue circles are calculated with interpolated energy-angular distributions. The ratios of damage energies computed with interpolated energy-angular distributions to the linear interpolated damage energies are show in the lower sub-plot. The ratios are close to unity. Therefore, the direct interpolation of energy-angular-integrated damage energies can give the same results as the damage computed with standard interpolated energy-angular distributions. It is shown that the interpolation of energy-angular distributions between two incident energies is not necessary for the calculation of DPA cross sections. By consequence, the computation of DPA cross sections can be largely

simplified using the direct interpolation of energy-angular-integrated damage energies.

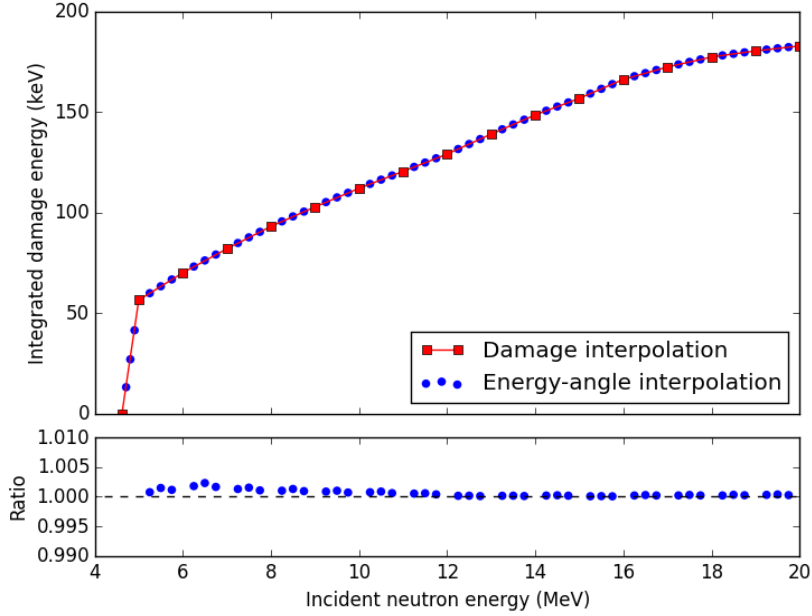

Figure 11. Energy-angular-integrated damage energy. The red square points are computed with energy-angular distributions given in the ENDF, the red lines are linear interpolation between two incident energies, the blue circles are calculated with interpolated energy-angular distributions. The lower figure plots the ratio of blues circles to the interpolated damage energies.

3.4 Role of high-order Legendre Polynomials

Figure 12 shows the neutron continuum inelastic scattering DPA cross sections of $^{56}$Fe computed with different maximum Legendre (Lmax) polynomials. Lmax = 0 is equivalent to the isotropic angular distribution. The DPA cross sections calculated under the assumption of the isotropic angular distribution is larger than those computed with all orders of Legendre polynomials. This is due to the forward oriented distribution of the emitted neutron at high incident energies, while Eq. (30) indicates that the recoil energy decreases with the cosine of the emission angle. The overestimation of DPA cross sections with Lmax = 0 corresponds with the results shown in Refs. [2], [4], [5].

Results in Figure 12 show that the first order Legendre polynomial (L1) can well reproduce the DPA cross section at incident energy lower than 20 MeV, while up to 12$^{th}$ order Legendre polynomials are required to describe the anisotropic angular distribution. Comparing the DPA cross section calculated with all Legendre polynomials in JEFF-3.1.1, the maximum deviation of the DPA cross section calculated with [L0 (isotropic) + L1] is +0.4%.

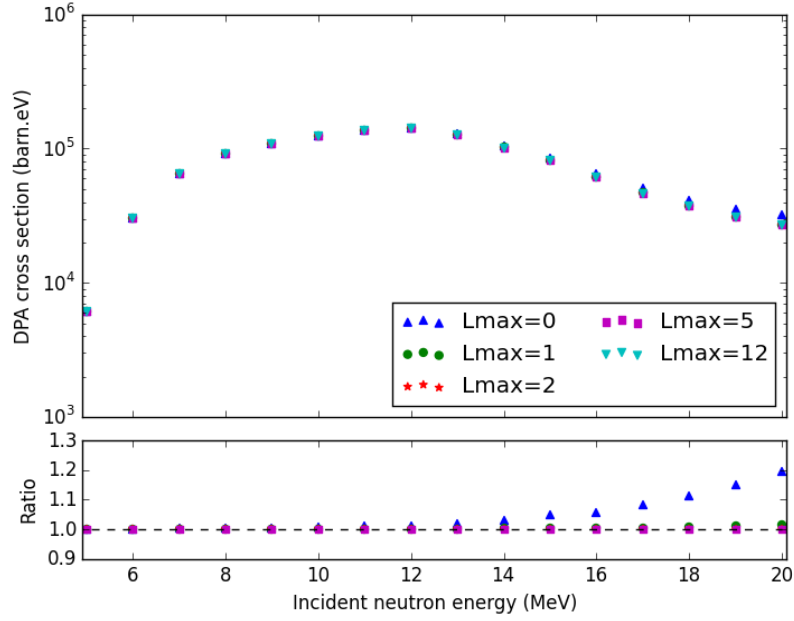

Figure 12. Neutron continuum inelastic scattering DPA cross sections of $^{56}$Fe performed with different maximum Legendre (Lmax) polynomials and the corresponding ratios to the DPA with full order Legendre polynomials.

## 4. Conclusions

The continuum inelastic scattering DPA cross section is computed with the double integral of the energy-angular distribution, which is recommended to be given in the Laboratory (Lab) frame. The routine of DPA calculation used in NJOY is the double integral over the emission angle and the secondary energy in this frame. However, the double-differential cross sections are often given in the Center-of-Mass (CM) frame. Three methods can be applied to perform the DPA calculations with data provided in the CM frame, including the transformation of data from the CM to the Lab frames (used in NJOY), the change of variables in double integrals, and the direct calculation. The first method increases the computation burden because the interpolation is required for each emission. For each given secondary energy, an angular integration should be performed. Moreover, additional error is introduced due to the loss of information during the transformation. The second one avoids the interpolations and additional error. Due to the Jacobian in the integrand, this method converges more slowly than the calculation with nuclear data given in the Lab frame. The last method is proposed and recommended in the present work for the energy-angular distributions provided in the CM frame. The direct calculation with the double-differential data in the CM frame is as simple as the method of DPA calculations with energy-angular distributions given in the Lab frame.

The DPA cross sections are computed with the double integration of the energy-angular distribution. The integrand of angular integration is the product of damage energy and energy-angular distribution described by Lmax orders Legendre polynomials for each secondary energy. The integrand is thus the product of a (Lmax+1)-order polynomial times a smooth non-polynomial function. Consequently, more than 1+[(Lmax+1)/2] points in the GLQ are required to ensure the convergence of numerical angular integration. For neutron continuum inelastic scattering of $^{56}$Fe in JEFF-3.1.1, more than 7 points are required to compute the angular integration. Numerical results show that 8-point GLQ is enough for the angular integration to compute DPA cross sections, while 64-point GLQ is used in NJOY for the energy-angular distribution given in ENDF MF6.

During the calculation of DPA cross sections, the integration over the secondary energy is performed with the trapezoidal integration with the energy grid given in ENDF. The integrand of the integration over the secondary energy is not a linear function. The computations with the 5 and 50 evenly inserted points in each interval of original energy are performed to verify the convergence of integration. Results show that the ENDF energy grid-based trapezoidal integration can calculate accurately the DPA cross sections due to the fine energy grid provided in ENDF-6 nuclear data files, at least for $^{56}$Fe.

Because the energy-angular distributions are only tabulated for several incident energies, the DPA cross sections between two neighbor incident energies cannot be directly computed with the double integration over the energy-angular distribution. The present work computes the energy-angular distributions between two incident energies using the standard ENDF interpolation method. The numerical results show that the directly interpolated energy-angular-integrated damage energies corresponds well with those computed with the standard interpolated energy-angular distributions. By consequence, the direct interpolation of energy-angular-integrated damage energies can be used to calculate DPA cross sections.

The high-order Legendre polynomials are of importance to describe the anisotropic angular distributions for neutron slowing down computation (and transport) problems for instance. However, the DPA cross section is an angle-integrated quantity. The high-order Legendre polynomials are less important for DPA calculations than for neutron flux and PKA spectra calculations. The DPA cross sections under the assumption of isotropic angular distribution are higher than those calculated with the anisotropic emission angle due to the forward oriented angular distribution of the scatterings, while the damage energy decreases with the cosine of the emission angle. Nevertheless, the addition of the first-order Legendre polynomial allows to compute the DPA cross sections within 0.4% deviation, while 12 orders are required to describe the anisotropy of angular distribution.